\documentclass{aa}
\usepackage{txfonts}
\usepackage{natbib}
\bibpunct{(}{)}{;}{a}{}{,} % to follow the A&A style
\bibliography{apj} % style apj.bst
\usepackage{graphicx}
\begin{document}
\title{Simultaneous radio-interferometric and high-energy TeV observations of the
       $\gamma$-ray blazar Mkn~421}
\author{P. Charlot\inst{1}
 \and  D. C. Gabuzda\inst{2}
 \and  H. Sol\inst{3}
 \and  B. Degrange\inst{4}
 \and  F. Piron\inst{5}
}
\offprints{Patrick Charlot,\\ \email{charlot@obs.u-bordeaux1.fr}} %%
\institute{Observatoire de Bordeaux (OASU) -- CNRS/UMR~5804,
           B.P.~89, 33270 Floirac, France
  \and     University College Cork -- Physics Department, Cork, Ireland
  \and     Observatoire de Paris -- CNRS/UMR~8631, 92195 Meudon Cedex, France
  \and     Ecole Polytechnique -- Laboratoire Leprince-Ringuet, 91128 Palaiseau, France
  \and     Universit\'e de Montpellier II, LPTA -- CNRS/UMR 5207,
           Place Bataillon, 34095~Montpellier Cedex~5, France
           }
\date{Received August 22, 2005/ Accepted April 21, 2006}
\abstract{The TeV-emitting BL~Lac object Mkn~421 was observed with
very long baseline interferometry (VLBI) at three closely-spaced
epochs one-month apart in March--April 1998. The source was also
monitored at very-high $\gamma$-ray energies (TeV~measurements)
during the same period in an attempt to search for correlations
between TeV variability and the evolution of the radio morphology on
parsec scales. While the VLBI maps show no temporal changes in the
Mkn~421 VLBI jet, there is strong evidence of complex variability in
both the total and polarized fluxes of the VLBI core of Mkn~421 and
in its spectrum over the two-month span of our data. The high-energy
measurements indicate that the overall TeV activity of the source
was rising during this period, with a $\gamma$-ray flare detected
just three days prior to our second VLBI observing run. Although no
firm correlation can be established, our data suggest that the two
phenomena (TeV activity and VLBI core variability) are connected,
with the VLBI core at 22~GHz being the self-absorbed radio
counterpart of synchrotron self-Compton (SSC) emission at high
energies. Based on the size of the VLBI core, we could derive an
upper limit of 0.1~pc ($3\times 10^{17}$~cm) for the projected size
of the SSC zone. This determination is the first model-free estimate
of the size of the $\gamma$-ray emitting region in a blazar.

\keywords{Galaxies: active -- Galaxies: jets -- BL~Lacertae objects:
individual: Mkn~421 -- Radio continuum: galaxies -- Techniques: high
angular resolution -- Gamma rays: observations}

}

\titlerunning{Simultaneous VLBI and TeV observations of the blazar Mkn~421}

\maketitle
%
%________________________________________________________________

\section{Introduction}

Among active galactic nuclei (AGN), BL~Lac objects form a class of
sources characterized by high radio and optical variability,
dominance of continuum over line emission, and strong and variable
polarization. Such extreme properties have already suggested long
ago that substantial relativistic beaming most probably occurs in
this type of source \citep{br78,bk79}. This idea was confirmed by
direct detection of apparent superluminal motion in the radio jets
of many BL~Lac objects based on high-resolution imaging with the
very long baseline interferometry (VLBI) technique
\citep{pm82,wsj88,gwr89,msb90,gpc00}. Building on these findings,
AGN unification schemes have emerged, all basically describing
BL~Lac objects as radio-loud AGN with relativistic jets pointing
within a small angle towards the observer
\citep{up95,fgb95,g95,kpl96,gm98,gcf98,fmc98,scu00,bl01}.

In the past decade, more evidence of relativistic beaming came from
high-energy X-ray and $\gamma$-ray data, which revealed intense
fluxes and strong variability up to the TeV level, especially in the
two nearby BL~Lac objects Mkn~421 and Mkn~501 \citep{pac92, qab96}.
In the case of Mkn~421, very high relativistic Doppler factors
(between 10 and 15) are demanded in order to reproduce the dramatic
rapid flares that have been observed on timescales as short as
30~min \citep{gab96,c97,cfr98,c04,aaa05}. Similar or even higher
Doppler factors (in the range 20--50) are also obtained when fitting
basic synchrotron self-Compton models to the broadband spectrum of
Mkn~421 \citep{mft99,tkm00,ksk01,kmk03,ksk03}.

The VLBI observations have sought evidence of superluminal motion in
\object{Mkn 421} ever since the early 1980's. A first series of VLBI
maps at 5~GHz reported apparent motions of about~2c and an angle to
the line of sight of $34^{\circ}$ for the parsec-scale jet
\citep{bel81,b84,zb90}. On the other hand, the source was found to
be unresolved at high frequency with a core size of
0.15~milliarcsecond (mas) at 22~GHz \citep{zb91}. In the 1990's, the
improved performances of the VLBI technique permitted the detection
of a weak one-sided jet \citep{pwx95,xrp95,emu98,kvz98,gfv99}. This
jet shows wiggles starting at about 5~mas from the core, as well as
strong distortions at a distance of 20~mas from the core. Based on
these data, \citet{gfv99} derived a viewing angle smaller than
$30^{\circ}$ and an apparent jet speed between $\sim$0.8c and 1c.
While faster speeds ($\sim$2c) were reported, invoking possible
earlier misidentification of several rapidly-evolving VLBI
components \citep{m96,m99}, another analysis based on a dense time
coverage (15~observing epochs over 3~years) confirmed the existence
of only subluminal apparent motion ($\le 0.3$c), therefore implying
a very small viewing angle to the line of sight of $0.4^{\circ}$ for
the VLBI jet \citep{puw99}. This analysis has recently been refined
with an extended data span (28~epochs over 8~years), leading to a
revised apparent speed of only $0.1\pm0.02$c for the fastest VLBI
jet component \citep{pe05}. While this value is smaller than the
value found by \citet{klh04}, who reported an average component
speed of $0.4$c from 15~GHz monitoring over 6~years, the two results
do rule out the existence of superluminal motion in the Mkn~421 jet.
Interestingly, Mkn~421 is not a unique case and such low apparent
speeds have been found in the other TeV blazars as well
\citep{pe04}.

The apparent inconsistency between the low degree of relativistic
beaming derived from the VLBI observations of TeV~blazars and the
high value predicted by the theory may be explained if the Doppler
factor decreases along the jet as a result of either jet curvature
or jet deceleration \citep{gk03,gtc05}. It is also possible that the
measured VLBI jet speed does not correspond to the speed of the
actual underlying jet but instead to the speed of a perturbed
pattern along the jet \citep{z97}. In this case, the radio core may
still be efficiently boosted although there would be no evidence for
relativistic beaming from apparent motions in the VLBI jet.
Nevertheless, one would expect correlations between radio core
properties and high-energy events if this hypothesis is correct. One
indication of the existence of these correlations comes from the
apparent connection between the epoch of emergence of new VLBI
components and the occurrence of strong X-ray and $\gamma$-ray
flares in several $\gamma$-ray emitting AGN
\citep{wuz93,prk95,uwl97,okk98,bwk98,b98,kko98,w99,mmm00}.
Furthermore, it appears that the $\gamma$-ray flares detected by the
EGRET detector onboard the Compton Gamma-Ray Observatory seem to
occur during the rising phase of high-frequency radio outbursts,
again suggesting a possible connection between radio and high-energy
properties \citep{vtl96,vt96,lvt00}. This is especially true for
Mkn~421, which occasionally shows multi-spectral flares (from radio
to TeV energies), as reported by \citet{ksk03}. Searching for
additional clues along these lines with new multi-frequency data is
important for investigating whether TeV blazars are indeed
strongly-beamed sources, as presumed so far.

This paper reports multi-frequency VLBI maps of Mkn~421 obtained at
three closely-spaced epochs (one month apart) in~1998. These
observations were arranged during regular monitoring of the source
at very-high $\gamma$-ray energy (TeV level) by the CAT Cherenkov
imaging telescope. In the following sections, we present results of
the VLBI and TeV observations, discuss the variability of the
source, and investigate possible connections between VLBI properties
and TeV activity. We also derive an upper limit for the projected
size of the $\gamma$-ray emission region based on the size of the
compact radio core as measured from our high-resolution VLBI data.

\section{Observations and data reduction}

\subsection{VLBI observations}

Multi-frequency (5.0~GHz, 8.4~GHz, 15.4~GHz, 22.2~GHz)
polarization-sensitive VLBI observations of Mkn~421 were carried out
at three epochs on March~4, March~28, and April~26, 1998, using the
Very Long Baseline Array (VLBA) telescope \citep{nbc94} of the
National Radio Astronomy Observatory (NRAO)\footnote{NRAO is
operated by Associated Universities, Inc., under cooperative
agreement with the National Science Foundation.}. Dual-polarization
(LCP and RCP) and two intermediate frequency channels (IFs), each
8~MHz wide, were recorded simultaneously at each band with scans of
3~min at 5~GHz and 8~GHz, 5~min at 15~GHz, and 6~min at 22~GHz.
Mkn~421 was observed alternately with another TeV-emitting source
(Mkn~501) at each of the four frequencies, along with interleaved
calibrators for amplitude and polarization calibration. The total
observing time for Mkn~421 was roughly 0.7~hr at 5~GHz and 8~GHz,
1.2~hr at 15~GHz, and 1.4~hr at 22~GHz.

The raw data were correlated with the VLBA correlator at the Array
Operations Center in Socorro, New Mexico, and calibrated with the
NRAO Astronomical Image Processing System in the standard way
\citep{u00}. The initial amplitude calibration for each of the IFs
and polarizations was accomplished using system temperature
measurements taken during the observations and NRAO-supplied gain
curves. The instrumental polarizations for each antenna were
determined from observations of the unpolarized source OQ208
(1404$+$286), while the absolute polarization position angles were
calibrated using observations of the compact polarized source
1823$+$568. For each epoch and frequency, the initial polarization
position angles have been rotated so that those for 1823$+$568 match
the integrated values reported in the University of Michigan Radio
Astronomy Observatory data base\footnote{see
http://www.astro.lsa.umich.edu/obs/radiotel/umrao.html} for our
observing epochs. The rotation values derived with this scheme
(equivalent to the R--L phase differences of the reference antenna)
were found to agree within $2\degr$ for the data of March~4, 1998,
and April~26, 1998, at all four frequencies. This is expected if the
R--L phase differences of the receivers are stable over this
timescale; in fact, the stability of this phase difference has been
demonstrated for the VLBA antennas on timescales as long as two
years \citep{rcg01}. It was intriguing, though, that the rotation
values derived for the data of the intermediate epoch (March~28,
1998) were inconsistent with these two sets of values. A closer look
at our calibrator data for March~28, 1998, revealed afterwards that
these data are indeed suspicious, probably because the polarization
properties of 1823$+$568 varied over the 12-hour span of our
observations. Therefore, we calibrated the absolute orientation of
the polarized position angles for the data of March~28, 1998, using
the mean of the rotation values determined for the other two epochs.

The calibrated visibility data of Mkn~421 for each frequency band
were processed using the Brandeis software to produce hybrid maps of
the distribution of the total intensity~$I$ and linear
polarization~$P$\footnote{$P=p{\rm e}^{2{\rm i}\chi}=mI{\rm
e}^{2{\rm i}\chi}$, where $p=mI$ is the polarized intensity, $m$ is
the fractional linear polarization, and $\chi$ is the position angle
of the electric vector on the sky.}. Maps of the linear polarization
were made by referencing the calibrated cross-hand fringes to the
parallel-hand fringes using the antenna gains determined in the
hybrid-mapping procedure, then Fourier-transforming the cross-hand
fringes, and CLEANing. One by-product of this procedure is to
register the $I$ and $P$ maps to within a small fraction of the
beamwidth, so that corresponding $I$ and $P$ images may be directly
superimposed. Subsequent to imaging, circular Gaussian models were
fitted to the self-calibrated visibility data using a non-linear
least-squares algorithm implemented in the Brandeis software. Such
models are useful for estimating the spectrum of the individual VLBI
components and for tracking temporal changes in the source
structure, as discussed below.

\subsection{TeV measurements}

Very-high $\gamma$-ray energy measurements on Mkn~421 were obtained
using the ground-based Cherenkov telescope CAT (``Cherenkov Array at
Th\'emis'') located in the French Pyr\'en\'ees. This telescope,
which has a reflector area of 17.8~${\rm m}^{2}$ and a very
high-definition camera with 546~pixels of $0.1^{\circ}\times
0.1^{\circ}$, detects the Cherenkov light emitted by secondary
particles resulting from cascades of primary high-energy
$\gamma$~rays and cosmic rays in the atmosphere \citep{bbb98}. In
this technique, air showers induced by $\gamma$~rays are efficiently
separated from the background of cosmic-ray showers by the shape and
direction of their images \citep{ldp98}. The energy threshold of
about $250$~GeV at zenith is achieved by using fast photo-tubes and
fast coincidence electronics, with an integration time of 12~ns
matching the short-lived Cherenkov flashes.

Mkn~421 was the first extragalactic source detected at the TeV level
\citep{pac92}, soon after its identification as a $\gamma$-ray
source by EGRET in 1992 \citep{lbc92}. Monitoring by CAT was
conducted from December~1996 (when this telescope began operating)
until June~2003 (when it ceased operations). During this period,
Mkn~421 was observed each year from November to June (when the
source is visible during night time at the CAT site) for about two
weeks per month (centered on the new Moon) within technical and
weather constraints. A detailed description of the $\gamma$-ray
signal extraction and data analysis is given in \citet{pdp01}. Light
curves have been produced for the integral flux above 250~GeV by
assuming a differential $\gamma$-ray spectrum index of $-2.9$
representative of all available spectral measurements for Mkn~421.

\section{Observational results}

\subsection{VLBI morphology and variability}

\begin{figure*}
\centering
\includegraphics[origin=br,bb=40 130 555 750,clip]{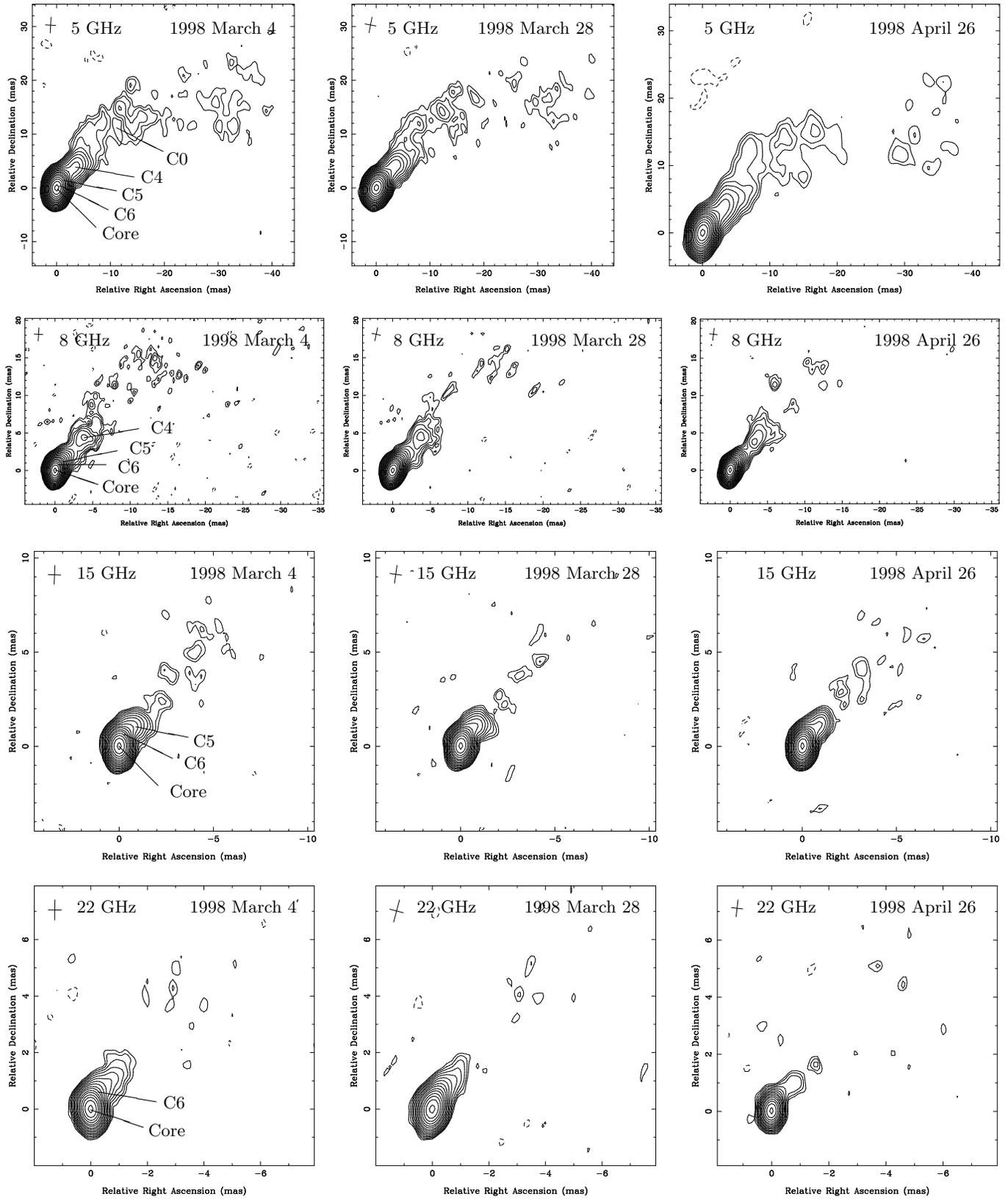}
\caption{Total-intensity maps of Mkn~421 at four frequencies (from
upper to lower panels: 5~GHz, 8.4~GHz, 15.4~GHz, and 22.2~GHz) and
three epochs (from left to right: March~4, March~28, and April~26,
1998). Image parameters are listed in Table~\ref{tab:maps}. The FWHM
Gaussian restoring beam applied to the images is shown as a cross in
the upper left-hand corner of each panel. Locations of the circular
Gaussian components fitted to the visibility data at each frequency,
as reported in Table~\ref{tab:models}, are marked in the left-hand
panels.} \label{fig:totmaps}
\end{figure*}

\begin{figure*}
\centering
\includegraphics[origin=br,bb=40 130 555 750,clip]{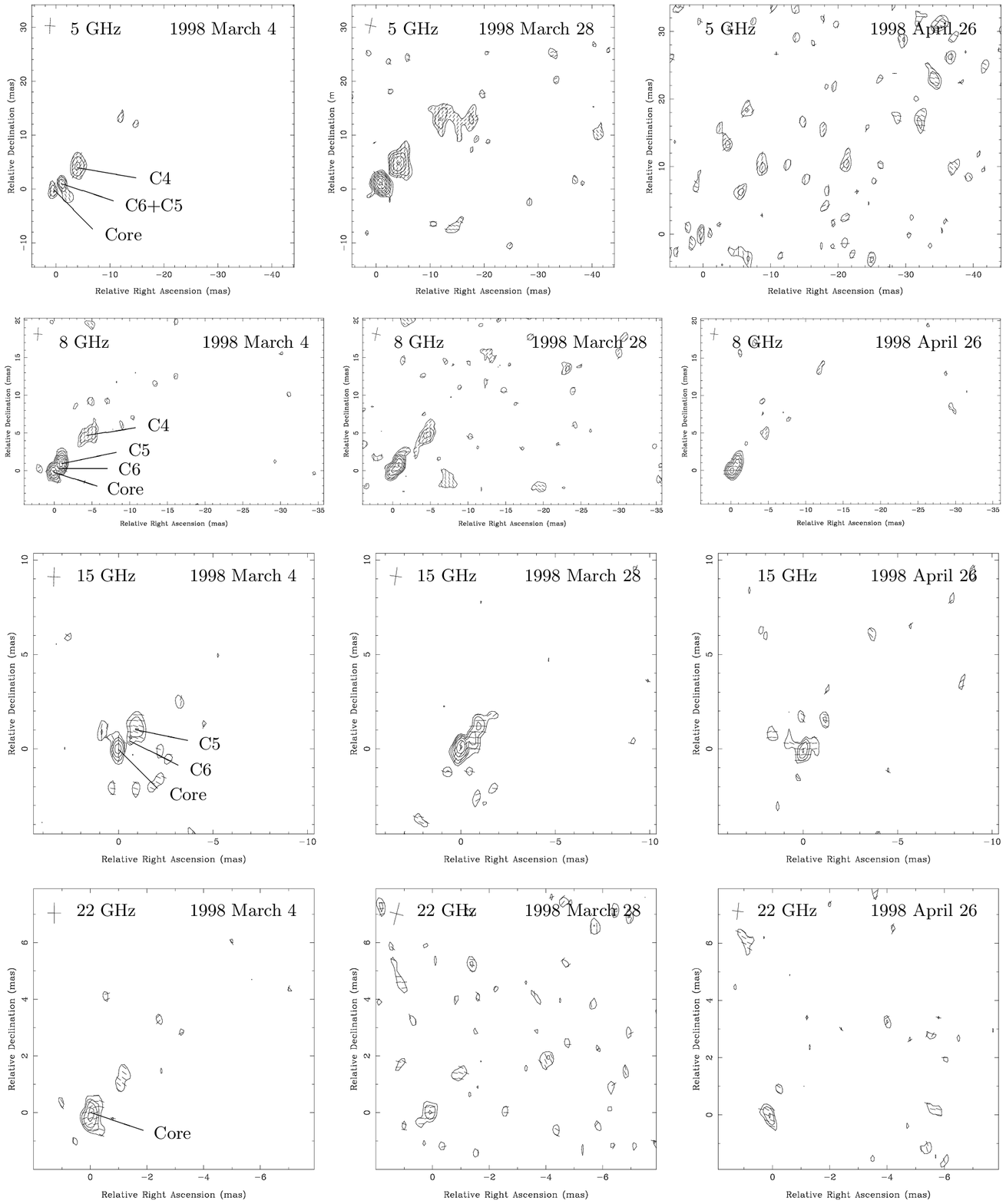}
\caption{Polarized-intensity maps of Mkn~421 at four frequencies
(from upper to lower panels: 5~GHz, 8.4~GHz, 15.4~GHz, and 22.2~GHz)
and three epochs (from left to right: March~4, March~28, and
April~26, 1998). Image parameters are listed in
Table~\ref{tab:maps}. The FWHM Gaussian restoring beam applied to
the images is shown as a cross in the upper left-hand corner of each
panel. Locations of the circular Gaussian components fitted to the
visibility data at each frequency, as reported in
Table~\ref{tab:models}, are marked in the left-hand panels.}
\label{fig:polmaps}
\end{figure*}

In order to provide an overview of our results in compact form, we
present our images of Mkn~421 at all four frequencies and all three
epochs in Figs.~\ref{fig:totmaps} and~\ref{fig:polmaps}. The image
parameters are listed in Table~\ref{tab:maps}. The total-intensity
morphology in our images is very similar to that observed in
previously published maps at these frequencies, such as in the large
series of images presented by \citet{puw99} and \citet{pe05}. We
identify the core and the three long-lived jet components C6, C5,
and C4 first detected by \citet{puw99}. The latest epoch of
\citet{puw99} was in December~1997, only three months before our
observations, and the cross-identification between our components
and theirs is straightforward.  Accordingly, we have retained this
naming convention for ease of comparison.  The additional 5~GHz
component labeled C0 is a combination of components C3, C2, and C1
from \citet{puw99}.


\begin{table*}
\begin{minipage}{\textwidth}
\caption{Parameters of the total-intensity and polarized-intensity
VLBI images of Mkn~421 plotted in Figs.~\ref{fig:totmaps}
\&~\ref{fig:polmaps}.} \label{tab:maps}
\renewcommand{\footnoterule}{}
\begin{tabular}{cccccrcccccc}
\hline\hline\noalign{\vskip 1.0mm}
&&&\multicolumn{3}{c}{Beam\footnote{The restoring beam is an
elliptical Gaussian with FWHM major axis $a$ and minor axis $b$,
with the major axis in position angle $\phi$ (measured from North
through East).}} &&\multicolumn{2}{c}{Total-intensity
maps}&&\multicolumn{2}{c}{Polarized-intensity maps}\\
\cline{4-6}\cline{8-9}\cline{11-12} \noalign{\vskip 1.0mm}
 Frequency & Epoch && a & b & $\phi$\ \ && Peak & Lowest contour\footnote{Successive contour levels are each a factor of the square
root of two higher.} && Peak & Lowest contour$^b$ \\
 (GHz)  &   && (mas) & (mas) & ($^{\circ} $)&& (mJy/beam) & \ \  ($\%$ of peak) && (mJy/beam) & \ \  ($\%$ of peak) \\
\noalign{\vskip 0.5mm} \hline \noalign{\vskip 0.5mm}
5.0  & 1 && 2.93 & 1.92 &  $-3.3$    && 357 & 0.18 && 3.98 & $35$ \\
     & 2 && 2.93 & 2.08 &  $-11.0$   && 324 & 0.25 && 5.95 & $12$ \\
     & 3 && 2.89 & 1.77 &  $-6.4$    && 317 & 0.25 && 1.46 & $55$\\\hline
8.4  & 1 && 1.76 & 1.15 &  $-3.3$    && 358 & 0.18 && 4.54 & $16$ \\
     & 2 && 1.79 & 1.17 &  $-12.0$   && 308 & 0.25 && 4.29 & $17$ \\
     & 3 && 1.63 & 1.07 &  $-6.9$    && 315 & 0.25 && 4.21 & $22$\\\hline
15.4 & 1 && 1.00 & 0.67 &  $-2.3$    && 325 & 0.35 && 6.07 & $28$ \\
     & 2 && 0.95 & 0.63 &  $-8.1$    && 296 & 0.50 && 7.92 & $17$ \\
     & 3 && 0.92 & 0.58 &  $-4.7$    && 295 & 0.35 && 4.72 & $34$\\\hline
22.2 & 1 && 0.72 & 0.46 &  $-0.8$    && 249 & 0.50 && 5.53 & $32$ \\
     & 2 && 0.82 & 0.49 &  $-14.8$   && 255 & 0.71 && 4.18 & $48$ \\
     & 3 && 0.69 & 0.40 &  $-5.5$    && 231 & 0.71 && 4.41 & $48$\\
\noalign{\vskip 0.5mm}\hline\hline
\end{tabular}
\end{minipage}
\end{table*}

As noted above, in order to derive a description of the source
structure that could be used to quantitatively compare our results
at the different frequencies and epochs, we obtained circular
Gaussian model fits to all the calibrated $I$ and $P$ visibility
data. The resulting Gaussian model components are presented in
Table~\ref{tab:models}. The errors reported in this table are our
best $1\sigma$ error estimates, as derived by examining the $\chi^2$
variations when forcing changes in the individual model components
(errors of $1\sigma$ correspond to an increase in the best-fit
$\chi^2$ by unity). In some cases these errors appear reasonable,
but in some other cases they clearly underestimate the true errors.
For example, reasonable errors in the separation from the core $r$
are probably no less than about 1/20th of the beam size (about 0.08,
0.06, 0.03, and 0.02~mas at 5, 8, 15, and 22~GHz, respectively),
which is larger than the errors reported for C6 and C5 in
Table~\ref{tab:models}, especially at the lowest frequencies (5 and
8~GHz).

\begin{table*}[p]
\centering
\begin{minipage}{\textwidth}
\renewcommand{\footnoterule}{}
\caption{Model-fitting results.} \label{tab:models}
\begin{tabular}{ccrrrrrrcrrrr}
\hline\hline\noalign{\vskip 1.0mm}
 Component\footnote{The components fitted to the
complex visibility data are of circular Gaussian form with FWHM axis
$d$, total intensity $I$, and polarized intensity~$p$. They are
separated from the (arbitrary) origin of the image by an amount $r$
in position angle $\theta$, which is the position angle (measured
from North through East) of a line joining the components with the
origin. The direction of the electric vector on the sky is given for
each component by its position angle $\chi$ (measured from North
through East).} & Epoch & $r$ \ \ \ & $\pm$ \ \ & $\theta$ \ \ &
$\pm$ \ &
$I$ \ \ \  & $\pm$ \  &  $p$  &$\pm$ \  &  $\chi$ \ \  & $\pm$ \  & $d$ \ \ \ \\
&&(mas)&&($\degr$)\ \  &&(mJy)&&(mJy)&&($\degr$)\ \  &&(mas)\\
\noalign{\vskip 0.5mm} \hline \noalign{\vskip 1.0mm}
\multicolumn{13}{c}{5 GHz}\\
\noalign{\vskip 1.0mm}
Core & 1 & ---  &  --- &  ---    & --- & 316.6 & 0.6 & 3.2     & 0.5 & $-36.5$ & 2.4 & 0.24  \\
     & 2 & ---  &  --- &  ---    & --- & 288.7 & 0.6 & 0.5     & 1.0 & $-30.0$ &10.0 & 0.23 \\
     & 3 & ---  &  --- &  ---    & --- & 260.4 & 0.8 & 1.6     & 0.5 & $-38.3$ &10.4 & 0.14    \\\hline
C6   & 1 & 0.86 & 0.01 & $-32.3$ & 0.7 &  57.0 & 0.6 & 4.5\footnote{Combinations of polarizations for C6$+$C5.}\hspace{-1.5mm} & 0.5 & $+44.0^b$\hspace{-1.5mm} & 2.4 & 0.32\\
     & 2 & 0.96 & 0.01 & $-34.0$ & 0.4 &  50.5 & 1.2 & 5.3$^b$\hspace{-1.5mm} & 0.6 & $+39.1^b$\hspace{-1.5mm} & 1.3 & 0.35 \\
     & 3 & 0.80 & 0.01 & $-35.2$ & 0.9 &  68.0 & 1.6 & ---     & --- &  ---    & --- & 0.05    \\\hline
C5   & 1 & 1.98 & 0.03 & $-41.4$ & 0.8 &  20.6 & 0.8 &         &     &         &     & 0.73 \\
     & 2 & 2.10 & 0.02 & $-41.6$ & 0.7 &  15.6 & 0.4 &         &     &         &     & 0.79 \\
     & 3 & 1.94 & 0.02 & $-40.6$ & 0.7 &  22.0 & 1.8 & ---     & --- &  ---    & --- & 0.57    \\\hline
C4   & 1 & 5.39 & 0.05 & $-40.0$ & 0.6 &  24.4 & 0.8 & 4.2     & 0.5 & $-34.5$ & 5.6 & 2.36\\
     & 2 & 5.36 & 0.08 & $-40.8$ & 0.7 &  23.7 & 0.6 & 4.0     & 0.8 & $-26.5$ & 2.8 & 2.47 \\
     & 3 & 5.28 & 0.08 & $-40.6$ & 0.8 &  25.5 & 0.8 & ---     & --- &  ---    & --- & 2.68  \\\hline
C0   & 1 &14.78 & 0.54 & $-44.0$ & 2.1 &  22.7 & 1.4 & ---     & --- &  ---    & --- &11.51\\
     & 2 &15.91 & 0.52 & $-43.6$ & 1.9 &  26.6 & 2.0 & ---     & --- &  ---    & --- &12.50 \\
     & 3 &15.57 & 0.49 & $-44.2$ & 1.8 &  22.6 & 3.2 & ---     & --- &  ---    & --- &10.93  \\
\noalign{\vskip 0.5mm} \hline\hline \noalign{\vskip 1.0mm}
\multicolumn{13}{c}{8 GHz}\\
\noalign{\vskip 1.0mm}
Core & 1 &  --- & ---  &  ---    & --- & 343.0 & 2.4 & 5.5  & 0.9 & $-48.5$ & 2.8 & 0.16 \\
     & 2 &  --- & ---  &  ---    & --- & 283.9 & 1.0 & 3.3  & 1.0 & $-62.9$ & 3.3 & 0.15 \\
     & 3 &  --- & ---  &  ---    & --- & 275.5 & 1.0 & 3.7  & 0.7 & $-75.4$ & 4.4 & 0.13 \\\hline
 C6  & 1 & 0.81 & 0.01 & $-24.4$ & 0.6 &  34.9 & 1.8 & 2.0  & 0.8 & $+61.8$ & 3.5 & 0.09 \\
     & 2 & 0.68 & 0.01 & $-30.2$ & 0.5 &  38.1 & 1.4 & 2.2  & 1.0 & $+58.6$ & 5.4 & 0.19 \\
     & 3 & 0.50 & 0.01 & $-29.1$ & 0.8 &  54.3 & 1.2 & 1.7  & 0.5 & $+67.3$ & 5.6 & 0.19 \\\hline
 C5  & 1 & 1.57 & 0.02 & $-41.9$ & 0.6 &  21.8 & 1.2 & 3.9  & 0.8 & $+55.1$ & 2.4 & 0.38 \\
     & 2 & 1.57 & 0.01 & $-38.7$ & 0.5 &  21.2 & 1.0 & 3.1  & 1.1 & $+55.7$ & 2.8 & 0.39 \\
     & 3 & 1.50 & 0.01 & $-38.0$ & 0.5 &  26.9 & 1.2 & 3.0  & 1.4 & $+53.8$ & 4.3 & 0.52 \\\hline
 C4  & 1 & 5.62 & 0.12 & $-41.4$ & 1.2 &  21.8 & 1.0 & 2.5  & 1.2 & $-28.5$ & 5.6 & 2.70 \\
     & 2 & 5.37 & 0.08 & $-40.9$ & 0.8 &  24.0 & 1.2 & 3.3  & 0.8 & $-28.6$ & 4.4 & 3.47 \\
     & 3 & 5.89 & 0.10 & $-41.3$ & 1.0 &  23.7 & 1.8 & 3.2  & 1.3 & $-32.1$ & 6.0 & 3.04 \\
\noalign{\vskip 0.5mm} \hline\hline \noalign{\vskip 1.0mm}
\multicolumn{13}{c}{15 GHz}\\
\noalign{\vskip 1.0mm}
Core & 1 & ---  & ---  &  ---    & --- & 334.0 & 3.6 & 6.0  & 1.5 & $-10.0$ & 4.2  & 0.16\\
     & 2 & ---  & ---  &  ---    & --- & 304.1 & 1.2 & 7.9  & 1.6 & $-35.2$ & 3.1  & 0.16\\
     & 3 & ---  & ---  &  ---    & --- & 306.1 & 1.2 & 5.3  & 0.5 & $-72.1$ & 4.4  & 0.16 \\\hline
 C6  & 1 & 0.78 & 0.02 & $-22.2$ & 2.9 &  31.3 & 3.0 & 1.2  & 2.1 & $+60.0$ & 13.6 & 0.10\\
     & 2 & 0.65 & 0.01 & $-21.8$ & 0.9 &  17.6 & 1.8 & 3.2  & 1.2 & $+75.8$ & 5.2  & 0.14\\
     & 3 & 0.55 & 0.01 & $-23.1$ & 1.1 &  19.2 & 3.0 & ---  & --- &   ---   & ---  & 0.10 \\\hline
 C5  & 1 & 1.49 & 0.06 & $-40.3$ & 2.3 &  13.9 & 2.0 & 2.5  & 1.8 & $+66.0$ & 10.2 & 0.63\\
     & 2 & 1.37 & 0.02 & $-37.6$ & 0.6 &  26.8 & 1.2 & 2.7  & 1.9 & $+97.0$ &  6.7 & 0.66\\
     & 3 & 1.36 & 0.03 & $-38.0$ & 1.4 &  20.7 & 6.2 & ---  & --- &   ---   & ---  & 0.63 \\
\noalign{\vskip 0.5mm} \hline\hline \noalign{\vskip 1.0mm}
\multicolumn{13}{c}{22 GHz}\\
\noalign{\vskip 1.0mm}
Core & 1 & ---  & ---  &  ---    & --- & 280.5 & 4.6 & 5.5  & 1.4 & $-60.0$ & 6.6 & 0.19 \\
     & 2 & ---  & ---  &  ---    & --- & 276.6 & 2.0 & 5.3  & 1.9 & $-75.0$ & 6.7 & 0.16\\
     & 3 & ---  & ---  &  ---    & --- & 242.3 & 2.4 & 4.5  & 1.2 & $-116.9$& 6.7 & 0.12\\\hline
 C6  & 1 & 0.74 & 0.02 & $-22.2$ & 1.7 &  30.4 & 1.0 & ---  & --- &   ---   & --- & 0.66 \\
     & 2 & 0.85 & 0.04 & $-26.6$ & 3.0 &  27.5 & 1.8 & ---  & --- &   ---   & --- & 0.63\\
     & 3 & 0.72 & 0.05 & $-20.8$ & 6.7 &  23.0 & 2.2 & ---  & --- &   ---   & --- & 0.81\\
\noalign{\vskip 0.5mm} \hline\hline \noalign{\vskip 1.0mm}
\end{tabular}
\end{minipage}
\end{table*}

\citet{pe05} conclude that the jet components C4, C5, and C6 were
present over the entire time range spanned by their observations
(from 1994 to 2002) and that they showed at most subluminal speeds
of expansion from the VLBI core (linear least-squares-fit speeds of
$0.089 \pm 0.066$c for C4, $0.095\pm 0.024$c for C5, and $0.029\pm
0.026$c for C6). Our images are consistent with these results, and
show no evidence of systematic motions in the locations of any of
these jet components over the two-month time interval covered by our
observations. Thus, our data exclude the possibility that the source
structure actually evolves much more rapidly than deduced so far,
but that the time intervals between previous VLBI images were too
small to track the jet-component evolution.

Our measurements in Table~\ref{tab:models} also do not show any
conclusive evidence of variability in the fluxes or polarizations of
the jet components over our three epochs, except for possible mild
changes at the level of 10--20 mJy in the total flux of the jet
component C6. In contrast, we have a strong indication of
variability in both the total flux and the polarized flux of the
VLBI core at all four frequencies over the roughly two months
covered by our three epochs. In particular, the total flux of the
core decreased by 30--70 mJy (corresponding to a change of 10--20\%)
between March~4 and April~26 at all four frequencies, and this
decrease was accompanied by a rotation of the core polarization
angle~$\chi$ by 30--60$^{\circ}$ at all frequencies except 5~GHz
(see Table~\ref{tab:models}). Although the core $\chi$ was
relatively constant at 5~GHz, the corresponding core polarized flux
was probably variable. These variations are discussed in more detail
below.

\subsection{TeV variability}

An extensive study of the CAT measurements acquired on Mkn~421
between 1996 and 2000 is found in \citet{pdp01}. As reported in this
paper, the level of the TeV emission for Mkn~421 changed
significantly from year to year. While at a low level in 1996--1997
(mean flux above~250~GeV: $\Phi_{>250{\rm GeV}} = 2.5 \pm 0.9 \times
10^{-11} {\rm cm^{-2} s^{-1}}$), it was more than twice as high the
next year, when our VLBI maps were obtained ($\Phi_{>250{\rm GeV}} =
6.0 \pm 0.6 \times 10^{-11} {\rm cm^{-2}s^{-1}}$ in 1997--1998).
Later on, the very-high-energy emission decreased to a level
comparable to that of 1996--1997 ($\Phi_{>250{\rm GeV}} = 2.7 \pm
0.9 \times 10^{-11} {\rm cm^{-2} s^{-1}}$ in 1998--1999), whereas
intense activity started again in January~2000.

The nightly-averaged integral flux above 250~GeV obtained by CAT
between December~1997 and May~1998 is plotted in
Fig.~\ref{mrk421_2000}. Also indicated in this figure are the epochs
of our three VLBI observing runs. The most striking features in the
light curve are the two $\gamma$-ray flares detected in January and
March~1998, the second of which occurred on March~25, just three
days prior to our second VLBI run (Fig.~\ref{mrk421_march}).
Additionally, the overall TeV activity seems to show a regular
increase from the first to third VLBI observing epochs, although
such a trend is difficult to establish because of the sparseness of
the TeV data, especially near the first VLBI run (due to poor
weather conditions). Most noticeably, the level of the TeV emission
near our third VLBI observing epoch (April~26) is significantly
higher than the mean TeV flux value for the campaign (materialized
by the horizontal dashed-line in Fig.~\ref{mrk421_2000}), while that
near our second VLBI run (March~28) is roughly at the same level.

\section{Discussion}

\subsection{Variability of the VLBI core}

As noted previously, both the total flux and the polarization
characteristics of the VLBI core were variable in the time spanned
by our observations. In this section, we first discuss the
variability in the total flux and its frequency dependence and then
examine the variability of the polarization flux and position angle
based on the model-fitting results in Table~\ref{tab:models}.

\begin{figure*}
\resizebox{\hsize}{!}{\includegraphics[angle=-90]{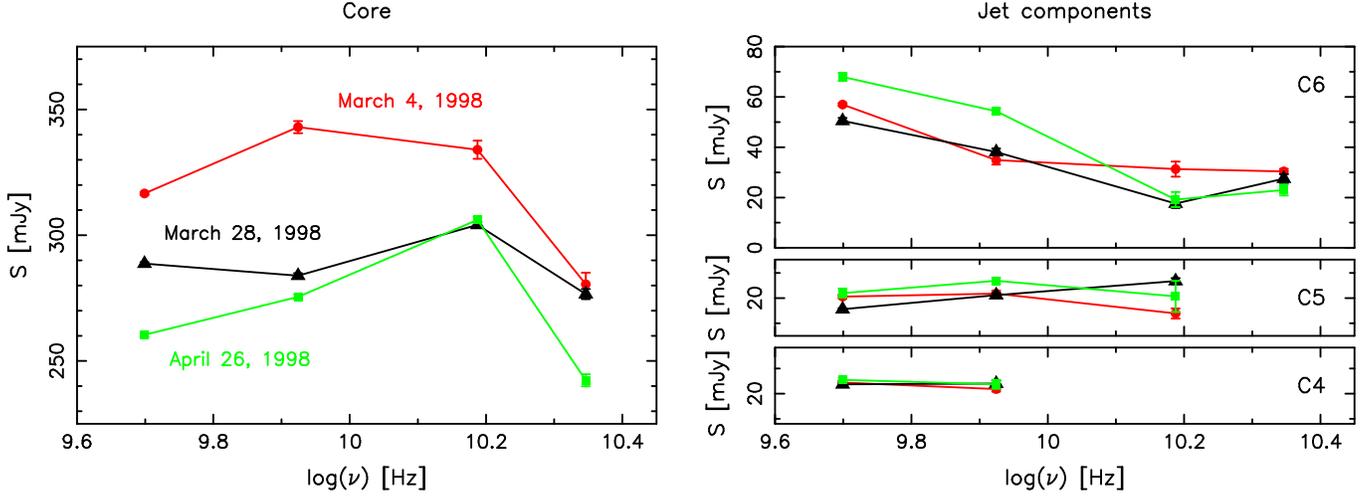}}
\caption{Spectra of the VLBI core and jet components C6, C5, and C4
for our three VLBI observing epochs (March~4, 1998, plotted in red,
March~28, 1998, plotted in black, and April~26, 1998 plotted in
green). The flux errors correspond to the formal $1\sigma$ errors
from the model-fitting results in Table~\ref{tab:models}. The same
flux scale is used in each panel.} \label{fig:spectra}
\end{figure*}

Figure~\ref{fig:spectra} presents the spectra of the core and the
jet components C6, C5, and C4. The core and C6 were detected at all
four frequencies, while C5 and C4 were detected only at the lowest
frequencies (5, 8, and 15~GHz for C5; 5 and 8~GHz for C4). The
fluxes for C5 and for C4 at the three different epochs all agree
within about 10~mJy or less and show no indication of variability.
Those for C6 show differences up to 20~mJy, especially at 5 and
8~GHz between our last two epochs. However, we cannot say for sure
whether these are real or due to systematic deviations in the flux
estimates. Overall, there is no convincing evidence of variability
in the jet-component fluxes. The spectrum of C6 is consistent with
this component being optically thin throughout the range of
frequencies we have observed (spectral index of about $-0.6$), as is
expected for jet emission. The spectra of C5 and C4 instead appear
to be flatter (spectral index of about 0.0 for C5 and $-0.1$ for
C4).

\begin{figure}
\resizebox{\hsize}{!}{\includegraphics[angle=0,bb=20 30 460
320]{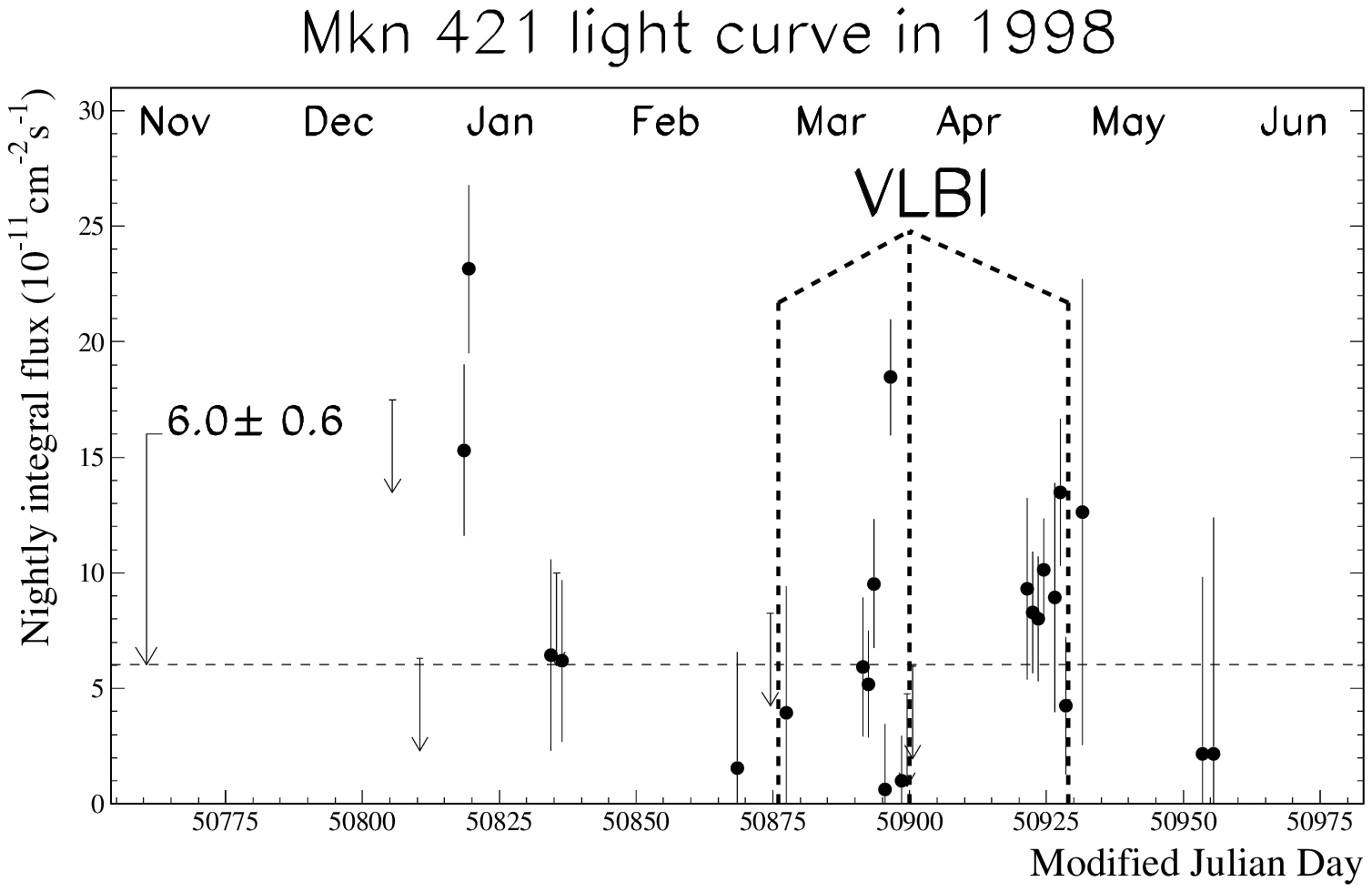}} \caption{Nightly-averaged integral flux above
$250\:\mathrm{GeV}$ for Mkn~421, as measured by CAT between
December~1997 and May~1998. Arrows stand for 2$\sigma$ upper-limits
when no signal was detected. The horizontal dashed line shows the
mean flux over the 5-month period. The epochs of our three VLBI runs
(corresponding to MJD=50876, 50900, and~50929) are indicated by
vertical thick dashed lines.} \label{mrk421_2000}
\end{figure}

In contrast, the core total flux is appreciably variable when
compared to the constancy of the flux densities of the jet
components C6, C5, and C4. As shown in Fig.~\ref{fig:spectra}, the
core flux density decreases at all four frequencies from our first
to our third epochs. The total changes are 56, 68, 28, and 38~mJy at
5, 8, 15, and 22~GHz, which are clearly significant, even if some of
the core flux errors are underestimated. The decrease in the core
flux is seen at each frequency, arguing against the possibility that
it is due to calibration errors, since the data for each frequency
were calibrated independently. In addition, if the flux density
decrease were associated with calibration errors, we might expect it
to be more prominent at the higher frequencies, while the opposite
is true. Figure~\ref{fig:spectra} also shows that the spectrum of
the core varies as the flux density decreases. The core spectrum at
our first epoch (March~4, 1998) has a maximum near 8~GHz and is
presumably self-absorbed at lower frequencies. The spectrum at our
second epoch three weeks later (March~28, 1998) is complex, but
roughly speaking has become flatter. The core spectrum at our third
epoch (April~26, 1998) is instead similar to the peaked spectrum
observed at our first epoch, though the flux density at 15~GHz is
somewhat higher than we would expect in this case. Thus, it seems
that the core spectrum first became flatter, then started to change
toward becoming peaked again, with an overall flux density level
appreciably below its initial level on March~4, 1998 (our first
epoch).

\begin{figure*}
\resizebox{\hsize}{!}{\includegraphics[angle=-90]{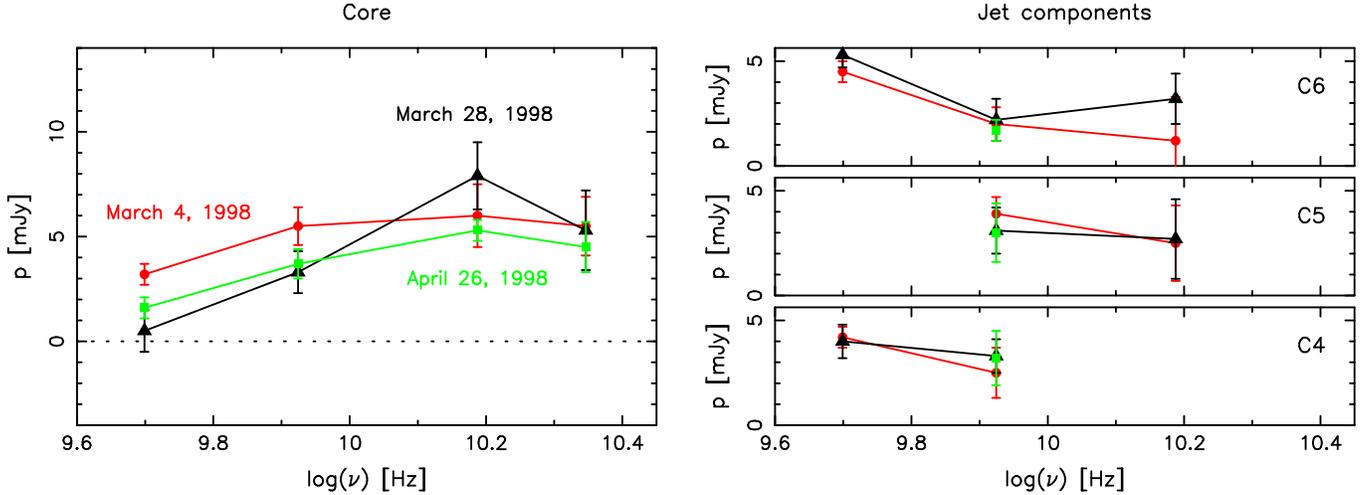}}
\caption{Polarized flux of the VLBI core and jet components C6, C5,
and C4 for our three VLBI observing epochs (March~4, 1998, plotted
in red, March~28, 1998, plotted in black, and April~26, 1998 plotted
in green). The flux errors correspond to the formal $1\sigma$ errors
from the model-fitting results in Table~\ref{tab:models}. The same
flux scale is used in each panel.} \label{fig:polarization}
\end{figure*}

There is also some evidence from our data that the polarized flux of
the core may have been variable at 5 and 8~GHz, though it is
difficult to be sure of this given the corresponding errors.
Figure~\ref{fig:polarization} plots the polarized fluxes of the core
and the jet components C6, C5, and C4 for the three successive
epochs. As in the previous case, the jet components were detected
only at the lowest frequencies (5, 8, and 15~GHz for C6; 8~and
15~GHz for C5; 5~and 8~GHz for C4). The core polarization was
clearly detected at 5~GHz at our first epoch (estimated flux of $3.2
\pm 0.5$~mJy), together with polarized flux associated with the jet
components C6 and C4; however, the core polarization was {\em not}
detected above the noise level at our second epoch (estimated flux
of $0.5 \pm 1.0$~mJy), though the polarizations from the jet
components C6 and C4 were detected and were equal to their values at
our first epoch to within the errors (see
Fig.~\ref{fig:polarization}). The core polarization was only weakly
detected at 5~GHz at our third epoch (estimated flux of $1.6 \pm
0.5$~mJy). Overall, the fractional core polarization over the three
epochs is 1\% at most, which is half of that measured by
\citet{ptz03} in a 5~GHz image from December 2000. The 8~GHz
measurements suggest that the core polarized flux also decreased at
this frequency between our first two epochs, but formally the
decrease does not exceed the error estimates (difference
of~$1.6\sigma$). At 15~and 22~GHz, the uncertainties are too large
to draw any firm conclusion about the core variability, even though
the core polarized flux appears to be somewhat weaker at the third
epoch than at the first epoch. The fractional core polarization at
these frequencies is about 2\% and is consistent with that measured
by \citet{mjm02} in a 22~GHz image from August~1997, but it is
smaller than the average value of 4.6\% reported by \citet{pe05}
from 22~GHz data acquired at four epochs in 2001--2002. Overall,
none of the jet components show signs of variability (see
Fig.~\ref{fig:polarization}).

\begin{figure}
\resizebox{\hsize}{!}{\includegraphics[angle=0, bb=20 30 460
320]{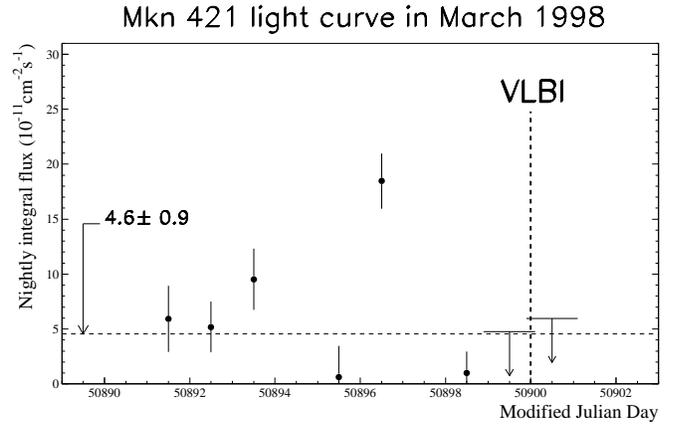}} \caption{Nightly-averaged integral flux above
$250\:\mathrm{GeV}$ for Mkn~421, as measured by CAT during the
10-day period prior to our second VLBI observation on March 28, 1998
(indicated by a vertical thick dash line at MJD=50900). Arrows stand
for 2$\sigma$ upper-limits when no signal was detected. The
horizontal dashed line shows the mean flux over this 10-day period.}
\label{mrk421_march}
\end{figure}

\begin{figure*}
\resizebox{\hsize}{!}{\includegraphics[angle=-90]{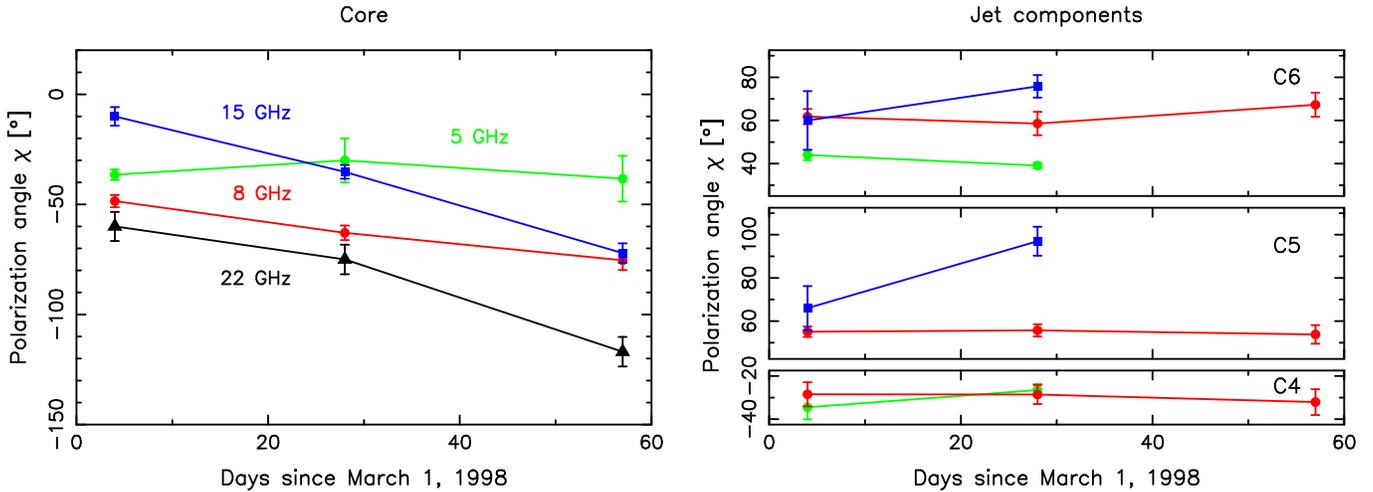}}
\caption{Polarization angle variations in the VLBI core and jet
components C6, C5, and C4 at 5~GHz (plotted in green), 8~GHz
(plotted in red), 15~GHz (plotted in blue), and 22~GHz (plotted in
black) over our three VLBI observing epochs (March 4, March 28, and
April 26, 1998). The polarization angle errors correspond to the
formal $1\sigma$ errors from the model-fitting results in
Table~\ref{tab:models}. The same $\chi$ scale is used in each
panel.} \label{fig:chi_rot}
\end{figure*}

Figure~\ref{fig:chi_rot} plots the variability of the polarization
position angle $\chi$ for the core and the jet components C6, C5,
and C4, based on the data in Table~\ref{tab:models}. As shown in
this figure, the core polarization position angle at 5~GHz is stable
over our three epochs and the orientation of the electric vectors at
this frequency is aligned with the direction of the inner VLBI jet
(about $-35^{\circ}$). This value for the polarization angle of the
core differs by $70^{\circ}$ from that measured by \citet{ptz03} in
December~2000, an indication of the existence of significant
variability on the long term even though none is detected over the
two-month span of our data. In contrast, the core polarization
position angles at 8,~15, and 22~GHz changed substantially in the
time between our first and third epochs, whereas no such changes
were found for the polarization angles of the jet components~C6, C5,
and C4. Again, this differs from the behavior observed in
2001--2002, where significant polarization angle variability was
detected for component C6 but not for the core \citep{pe05}. In all
cases, the rotations for the core exceed the $3\sigma$~errors in the
polarization angles; the most dramatic rotation is through more than
$60^{\circ}$ at 15~GHz, corresponding to about~$10\sigma$ (see
Table~\ref{tab:models}). The direction of the rotation is the same
at all three frequencies, with the $\chi$~values becoming more
negative and showing a roughly linear dependence on time. Since the
rotation is not larger at the lower frequencies, it cannot be
associated with variable Faraday rotation in the VLBI core. The
$\chi$~orientation at 15~GHz at our first epoch ($-10^{\circ}$) is
nearly aligned with the direction of the inner VLBI jet (about
$-25^{\circ}$), and the $\chi$~orientation at 22~GHz at our third
epoch ($-117^{\circ}$) is nearly perpendicular to this direction.
However, the other core $\chi$~orientations at 8, 15, and 22~GHz do
not bear clear relationships to the direction of the inner jet. The
observed core electric vectors at 15 and 22~GHz rotate nearly in
synchrony, so that there is a large ($\simeq 50^{\circ}$) and nearly
constant offset between the $\chi$ values at these two frequencies,
whose origin is unclear; as discussed above, this offset does not
appear to be associated with Faraday rotation in the VLBI core.

\begin{figure}
\resizebox{\hsize}{!}{\includegraphics[angle=-90]{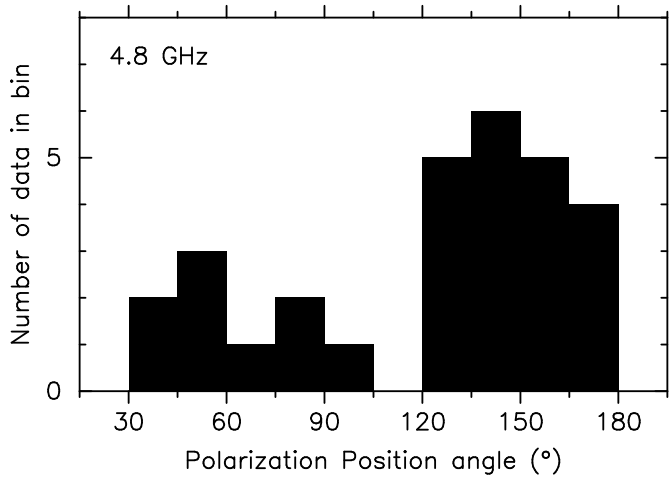}}
\caption{Distribution of the position angle for the total integrated
polarization at 4.8 GHz as observed by the University of Michigan
within 100 days of our mid-VLBI epoch. Note two perpendicular states
of polarization at about $\chi=150^{\circ}$ and $\chi=60^{\circ}$ in
these measurements.} \label{fig:mich_xi}
\end{figure}

The complex frequency dependence and variability of the core
polarization position angles may be due to the presence of
underlying components with different polarization properties within
the region encompassed by the VLBI core. If such subcomponents have
different polarization position angles and/or vary on different
timescales, a complex behavior arises naturally for the polarization
angles of the VLBI core when the individual subcomponents are
superimposed. Repeated measurements of the total integrated
polarization for Mkn~421, which indicate two predominating
polarization states over the time span of our observations
(Fig.~\ref{fig:mich_xi}), give support to this scenario. As seen in
Fig.~\ref{fig:mich_xi}, the two polarization states have electric
vectors that are either parallel to the VLBI jet
($\chi=150^{\circ}$) or perpendicular to the jet
($\chi=60^{\circ}$), hence corresponding to orthogonal magnetic
fields if assuming optically-thin synchrotron emission. In this
scenario, the dominance of either state in the integrated
polarization measurements would be the result of relative flux
variations occurring on short timescales for subcomponents within
the VLBI core. Alternately, the two polarization states may also
result from variations in the optical depth (thin or thick emission)
of such subcomponents if the underlying magnetic field is solely
perpendicular to the jet.

\subsection{VLBI properties and TeV activity}

Our VLBI data have revealed striking temporal changes over a few
weeks for the compact radio core in Mkn~421. To our knowledge, it is
the first time that such substantial variations on short timescales
have been observed by VLBI in a BL~Lac object (with the exception of
intraday variability). As noted previously, the complex temporal
evolution for the VLBI core of Mkn~421 occurred at a time when the
overall TeV $\gamma$-ray activity for Mkn~421 was rising
(Fig.~\ref{mrk421_2000}); furthermore, a TeV flare was detected just
three days prior to our second VLBI run (Fig.~\ref{mrk421_march}).
Both results qualitatively argue in favor of a link between radio
VLBI and TeV $\gamma$-ray properties.

Assuming that the $\gamma$-ray radiation is nested in the compact
VLBI core at 22~GHz (the highest available frequency), we can
directly constrain the size of the high-energy emission region
(projected on the plane of the sky) to be smaller than 0.15~mas (our
measured VLBI core FWHM at 22~GHz, see Table~\ref{tab:models}) or,
equivalently, 0.1~pc ($3\times 10^{17}$~cm) if adopting a distance
of 125~Mpc for Mkn~421 ($z=0.031$, $H_0=71$~km~s$^{-1}$~Mpc$^{-1}$).
This upper limit for the projected size of the $\gamma$-ray emitting
zone is consistent with typical estimates on the order of
$10^{16}$~cm derived from variability data \citep{k00} or from
broadband spectrum modeling \citep{ksk03}. While our determination
does not improve over such estimates, it is solely based on
high-resolution radio-interferometric observations and hence is
independent of any theoretical model.
%%1~mas~$=$~0.66~pc

\begin{figure}
\resizebox{\hsize}{!}{\includegraphics[angle=0,bb=10 27 535
385]{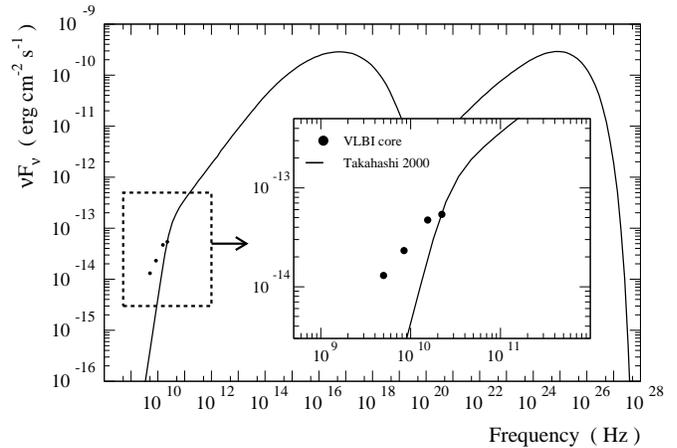}} \caption{Multi-frequency spectral energy
distribution from radio to TeV energies and VLBI core flux for
Mkn~421. The continuous line is a prediction based on the one-zone
synchrotron self-Compton model derived by \citet{tkm00} from the
data of the multi-wavelength campaign conducted in late April~1998,
while the dots represent our measured VLBI core fluxes at 5, 8, 15,
and 22~GHz on April 26, 1998 (see Table~\ref{tab:models}). The inset
shows an enlargement of the radio region.} \label{TakaSSC}
\end{figure}

In order to gain insight into the origin of the radiation for the
compact radio core in Mkn~421, we compared its flux with predictions
from high-energy emission models. Figure~\ref{TakaSSC} shows the
radio-to-TeV spectral energy distribution derived by fitting a
one-zone synchrotron self-Compton (SSC) model to the data of a
multi-frequency campaign on Mkn~421 that happened to coincide with
our third VLBI run \citep{tkm00}. Superimposed on this theoretical
curve, our measured VLBI core fluxes at 5, 8, 15, and 22~GHz (as
reported in Table~\ref{tab:models}) are also plotted. It is striking
that the flux of the VLBI core at the highest radio frequency
(22~GHz) is perfectly consistent with the prediction of the model,
whereas those at the lower radio frequencies (15, 8, and 5~GHz)
depart from the SSC~predictions significantly. This finding is a
strong indication that the compact core observed by VLBI at 22~GHz
is dominated by the radio counterpart of the X-ray and $\gamma$-ray
SSC radiation. Conversely, the VLBI core at 15~GHz and below must
encompass external radio-emitting regions located beyond the
$\gamma$-ray emitting zone, since the measured VLBI core fluxes at
these frequencies are higher than the predictions of the SSC~model.

The above scenario, which interprets the 22~GHz radio emission from
the VLBI core of Mkn~421 as the low-energy tail of a one-zone SSC
radiation, naturally implies a connection between radio variability
and TeV~activity. The existence of this connection is also supported
by single-dish radio monitoring which revealed that some radio
flares are correlated with TeV~flares, with a simultaneous increase
in the radio and high-energy fluxes when such events occur
\citep{ksk03}. In our case, however, we instead detect an overall
decrease in the radio flux of the VLBI core (Fig.~\ref{fig:spectra})
at a time when the very-high-energy $\gamma$-ray flux of the source
tends to increase (Fig.~\ref{mrk421_2000}). As noted by
\citet{ksk03}, self-absorption effects in the radio range can
produce anticorrelation between flux variations observed at
different energies within a SSC scenario. This may well be the case
for our VLBI and TeV data, since the four observed radio frequencies
are clearly within the self-absorbed part of the Mkn~421 spectrum
(Fig.~\ref{TakaSSC}). Under this assumption, the observed
anticorrelated radio and very-high-energy variations would fit
within the proposed SSC scenario.

\section{Conclusion}

Coordinated multi-frequency VLBI polarization and very-high-energy
TeV observations of the $\gamma$-ray blazar Mkn~421 have provided
evidence of variability in the compact VLBI core on a timescale of a
few weeks and at a time when the overall TeV activity of the source
was rising. Although we cannot be entirely sure about the connection
between the two phenomena, our measurements are consistent with a
scenario in which the VLBI core at 22~GHz represents the
self-absorbed part of the low-energy emission induced by a one-zone
synchrotron self-Compton model consistent with simultaneous
high-energy data. At the lower radio frequencies (15, 8, and 5~GHz),
the VLBI core encompasses additional emission unrelated to the SSC
phenomenon and most probably originating from regions outside the
$\gamma$-ray emitting zone. Based on these observations, we also
derived an upper limit of 0.1~pc ($3\times 10^{17}$~cm) for the
projected size of the $\gamma$-ray emitting region, in agreement
with previous estimates but based solely on our
radio-interferometric data and hence free of any theoretical model.

Further similar coordinated VLBI and TeV observations on Mkn~421
should be primarily targeted at improving the VLBI time coverage and
angular resolution. Observations a few days apart over several weeks
would be important to confirm whether the VLBI core variability and
TeV activity are indeed connected and whether there is any
anticorrelation between such variations. Increasing the VLBI angular
resolution, e.g. by observing at 43~GHz with a transatlantic network
or at 86~GHz with the global millimeter VLBI array, would also be
important to further constrain the location and size of the
$\gamma$-ray emission region in the framework of the proposed SSC
scenario. By providing an angular resolution improved by a factor of
up to~6, such high-frequency VLBI observations may perhaps even
permit a direct measurement of the size of the SSC zone, which is on
the order of $10^{16}$~cm according to present models. VLBI
monitoring of Mkn~421 should also be pursued on the long-term to
refine jet component proper motions and possibly obtain more clues
as to why there are no superluminal motions in this presumably
strongly-beamed source.

\begin{acknowledgements}
We thank T. Takahashi and J. Kataoka for providing the parameters of
their SSC model. This research has made use of data from the
University of Michigan Radio Astronomy Observatory, which has been
supported by the University of Michigan and the National Science
Foundation.
\end{acknowledgements}

\listofobjects


\begin{thebibliography}{}

\bibitem[Aharonian et al.(2005)]{aaa05}
Aharonian, F., Akhperjanian, A. G., Aye, K.-M., et al., 2005, A\&A, 437, 95

\bibitem[B\aa\aa th(1984)]{b84}
B\aa\aa th, L. B., 1984, in IAU Symposium 110: VLBI and Compact
Radio Sources, eds. R. Fanti, K. Kellermann, \& G. Setti (Dordrecht:
Reidel), 127

\bibitem[B\aa\aa th et al.(1981)]{bel81}
B\aa\aa th, L. B., Elgered, G., Lundqvist, G., et al., 1981, A\&A,
96, 316

\bibitem[Bai \& Lee(2001)]{bl01}
Bai, J. M., \& Lee, M. G., 2001, ApJ, 548, 244

\bibitem[Barrau et al.(1998)]{bbb98}
Barrau, A., Bazer-Bachi, R., Beyer, E., et al., 1998, NIM, A416, 278

%%\bibitem[Bennett et al.(2003)]{bhh03}
%%Bennett, C. L., Halpern, M., Hinshaw, G., et al., 2003, ApJS, 148, 1

\bibitem[Blandford \& K\"{o}nigl(1979)]{bk79}
Blandford, R. D., \& K\"{o}nigl, A., 1979, ApJ, 232, 34

\bibitem[Blandford \& Rees(1978)]{br78}
Blandford, R. D., \& Rees, M. J., 1978, in Proceedings of the
Pittsburgh Conference on BL~Lac Objects, ed. A. M. Wolfe
(Pittsburgh: University of Pittsburgh Press), 328

\bibitem[Bower(1998)]{b98}
Bower, G. C., 1998, in IAU Colloquium 164: Radio Emission from
Galactic and Extragalactic Compact Sources, eds. J.~A. Zensus, G. B.
Taylor, \& J. M. Wrobel, ASP Conf. Ser., Vol.~144, 41

\bibitem[Britzen et al.(1998)]{bwk98}
Britzen, S., Witzel, A., Krichbaum, T. P., Roland, J., \& Wagner,
S., 1998, in IAU Colloquium 164: Radio Emission from Galactic and
Extragalactic Compact Sources, eds. J.~A. Zensus, G. B. Taylor, \&
J. M. Wrobel, ASP Conf. Ser., Vol.~144, 43

\bibitem[Celotti et al.(1998)]{cfr98}
Celotti, A., Fabian, A. C., \& Rees, M. J., 1998, MNRAS, 293, 239

\bibitem[Coppi(1997)]{c97}
Coppi, P., 1997, in Relativistic Jets in AGNs, eds. M. Ostrowski, M.
Sikora, G. Madejski, \& M. Begelman (Cracow), 333

\bibitem[Cui(2004)]{c04}
Cui, W., 2004, ApJ, 605, 662

\bibitem[Edwards et al.(1998)]{emu98}
Edwards, P. G., Moellenbrock, G. A., Unwin, S. C., Wehrle, A. E.,
Weekes, T. C., 1998, in IAU Colloquium 164: Radio Emission from
Galactic and Extragalactic Compact Sources, eds. J. A. Zensus, G. B.
Taylor, \& J. M. Wrobel, ASP Conf. Ser., Vol. 144, 45

\bibitem[Falcke et al.(1995)]{fgb95}
Falcke, H., Gopal-Krishna, \& Biermann, P. L., 1995, A\&A, 298, 395

\bibitem[Fossati et al.(1998)]{fmc98}
Fossati, G., Maraschi, L., Celotti, A., Comastri, A., \& Ghisellini,
G., 1998, MNRAS, 299, 433

\bibitem[Gabuzda et al.(1989)]{gwr89}
Gabuzda, D. C., Wardle, J. F. C., \& Roberts, D. H., 1989, ApJ, 336,
L59

\bibitem[Gabuzda et al.(2000)]{gpc00}
Gabuzda, D. C., Pushkarev, A. B., \& Cawthorne, T. V., 2000, MNRAS,
319, 1109

\bibitem[Gaidos et al.(1996)]{gab96}
Gaidos, J. A., Akerlof, C. W., Biller, S. D., et al., 1996, Nature,
383, 319

\bibitem[Georganopoulos \& Kazanas(2003)]{gk03}
Georganopoulos, M., \& Kasanas, D., 2003, ApJ, 594, L27

\bibitem[Georganopoulos \& Marscher(1998)]{gm98}
Georganopoulos, M., \& Marscher, A. P., 1998, ApJ, 506, 621

\bibitem[Ghisellini et al.(1998)]{gcf98}
Ghisellini, G., Celotti, A., Fossati, G., Maraschi, L., \& Comastri,
A., 1998, MNRAS, 301, 451

\bibitem[Ghisellini et al.(2005)]{gtc05}
Ghisellini, G., Tavecchio, F., \& Chiaberge, M., 2005, A\&A, 432,
401

\bibitem[Giovannini et al.(1999)]{gfv99}
Giovannini, G., Feretti, L., Venturi, T., Cotton, W. D., \& Lara,
L., 1999, in BL~Lac Phenomenon, eds. L. O. Takalo, \& A.
Sillanp\"a\"a, ASP Conf. Ser., Vol. 159, 439

\bibitem[Gopal-Krishna(1995)]{g95}
Gopal-Krishna, 1995, JApAS, 16, 153

\bibitem[Kataoka (2000)]{k00}
Kataoka, J., 2000, Ph. D. Thesis, University of Tokyo

\bibitem[Katarzynski et al.(2003)]{ksk03}
Katarzynski, K., Sol, H., \& Kus, A., 2003, A\&A, 410, 101

\bibitem[Kellermann et al.(1998)]{kvz98}
Kellermann, K. I., Vermeulen, R. C., Zensus, J. A., \& Cohen, M. H.,
1998, AJ, 115, 1295

\bibitem[Kellermann et al.(2004)]{klh04}
Kellermann, K. I., Lister, M. L., Homan, D. C., et al., 2004, ApJ,
609, 539

\bibitem[Kollgaard et al.(1996)]{kpl96}
Kollgaard, R. I., Palma, C., Laurent-Muehleisen, S. A., \&
Feigelson, E. D., 1996, ApJ, 465, 115

\bibitem[Konopelko et al.(2003)]{kmk03}
Konopelko, A., Mastichiadis, A., Kirk, J., de Jager, O. C., \&
Stecker, F. W., 2003, ApJ, 597, 851

\bibitem[Krawczynski et al.(2001)]{ksk01}
Krawczynski, H., Sambruna, R., Kohnle, A., et al., 2001, ApJ, 559, 187

\bibitem[Krichbaum et al.(1998)]{kko98}
Krichbaum, T. P., Kraus, A., Otterbein, K., et al., 1998, in IAU
Colloquium 164: Radio Emission from Galactic and Extragalactic
Compact Sources, eds. J. A. Zensus, G. B. Taylor, \& J. M. Wrobel,
ASP Conf. Ser., Vol. 144, 37

\bibitem[L\"ahteenm\"aki et al.(2000)]{lvt00}
L\"ahteenm\"aki, A., Valtaoja, E., \& Tornikoski, M., 2000, in
Proceedings of The Fifth Compton Symposium, eds. M. L. McConnell,
\&. J. M. Ryan, AIP Conf. Proc., Vol. 510, 372

\bibitem[Le Bohec et al.(1998)]{ldp98}
Le Bohec, S., Degrange, B., Punch, M., et al., 1998, NIM, A416, 425

\bibitem[Lin et al.(1992)]{lbc92}
Lin, Y. C., Bertsch, D. L., Chiang, J., et al., 1992, ApJ, 401, L61

\bibitem[Maraschi et al.(1999)]{mft99}
Maraschi, L., Fossati, G., Tavecchio, F., et al., 1999, ApJ, 526, L81

\bibitem[Marscher(1996)]{m96}
Marscher, A. P., 1996, in Gamma-ray Emitting AGN, eds. J.~G. Kirk,
M. Camenzind, C. von Montigny, \& S. Wagner (Heidelberg), 97

\bibitem[Marscher(1999)]{m99}
Marscher, A. P., 1999, Astroparticle Physics, 11, 19

\bibitem[Marscher et al.(2002)]{mjm02}
Marscher, A. P., Jorstad, S. G., Mattox, J. R., \& Wehrle, A. E.,
2002, ApJ, 577, 85

\bibitem[Marchenko et al.(2000)]{mmm00}
Marchenko, S. G., Marscher, A. P., Mattox, J. R., et al., 2000, in
Proceedings of The Fifth Compton Symposium, eds. M.~L. McConnell,
\&. J. M. Ryan, AIP Conf. Proc., Vol. 510, 357

\bibitem[Mutel et al.(1990)]{msb90}
Mutel, R. L., Su, Bumei, Bucciferro, R. R., \& Phillips, R. B.,
1990, ApJ, 352, 81

\bibitem[Napier et al.(1994)]{nbc94}
Napier, P. J., Bagri, D. S., Clark, B. G., et al., 1994, Proc. IEEE,
82, 658

\bibitem[Otterbein et al.(1998)]{okk98}
Otterbein, K., Krichbaum, T. P., Kraus, A., et al., 1998, A\&A, 334,
489

%%\bibitem[Padovani(1992)]{p92}
%%Padovani, P., 1992, MNRAS 257, 404

\bibitem[Phillips \& Mutel(1982)]{pm82}
Phillips, R. B., \& Mutel, R. L., 1982, ApJ 257, L19

\bibitem[Piner \& Edwards(2004)]{pe04}
Piner, B. G., \& Edwards, P. G., 2004, ApJ, 600, 115

\bibitem[Piner \& Edwards(2005)]{pe05}
Piner, B. G., \& Edwards, P. G., 2005, ApJ, 622, 168

\bibitem[Piner et al.(1999)]{puw99}
Piner, B. G., Unwin, S. C., Wehrle, A. E., et al., 1999, ApJ, 525,
176

%%\bibitem[Piron(1998)]{p98}
%%Piron, F., for the CAT collaboration, 1998, 19th Texas Symposium on
%%Relativistic Astrophysics and Cosmology, Paris

%%\bibitem[Piron(1999)]{p99}
%%Piron, F., for the CAT collaboration, 1999, 26th International
%%Cosmic-Ray Conference (Salt-Lake City), 3, 326

\bibitem[Piron et al.(2001)]{pdp01}
Piron, F., Djannati-Ata\"\i, A., Punch, M., et al., 2001, A\&A, 374,
895

\bibitem[Pohl et al.(1995)]{prk95}
Pohl, M., Reich, W., Krichbaum, T. P., et al., 1995, A\&A, 303, 383

\bibitem[Polatidis et al.(1995)]{pwx95}
Polatidis, A. G., Wilkinson, P. N., Xu, W., et al., 1995, ApJS, 98,
1

\bibitem[Pollack et al.(2003)]{ptz03}
Pollack, L. K., Taylor, G. B., \& Zavala, R. T., 2003, ApJ, 589, 733

\bibitem[Punch et al.(1992)]{pac92}
Punch, M., Akerlof, C. W., Cawley, M. F., et al., 1992, Nature, 358,
477

\bibitem[Quinn et al. (1996)]{qab96}
Quinn, J., Akerlof, C. W., Biller, S., et al., 1996, ApJ, 456, L83

\bibitem[Reynolds et al.(2001)]{rcg01}
Reynolds, C, Cawthorne, T. V., \& Gabuzda, D. C., 2001, MNRAS, 327,
1071

\bibitem[Sambruna et al.(2000)]{scu00}
Sambruna, R. M., Chou, L. L., \& Urry, C. M., 2000, ApJ, 533, 650

\bibitem[Takahashi et al.(2000)]{tkm00}
Takahashi, T., Kataoka, J., \& Madejski, G., et al., 2000, ApJ, 542,
L105

\bibitem[Ulvestad(2000)]{u00}
Ulvestad, J., 2000, VLBA Scientific Memo No. 25, NRAO, Socorro

\bibitem[Unwin et al.(1997)]{uwl97}
Unwin, S., Wehrle, A. E., Lobanov, A. P., et al., 1997, ApJ, 480,
596

\bibitem[Urry \& Padovani(1995)]{up95}
Urry, C. M., \& Padovani, P., 1995, PASP, 107, 803

\bibitem[Valtaoja \& Ter\"asranta(1996)]{vt96}
Valtaoja, E., \& Ter\"asranta, H., 1996, A\&AS, 120, 491

\bibitem[Valtaoja et al.(1996)]{vtl96}
Valtaoja, E., Ter\"asranta, H, \& L\"ahteenm\"aki, A., 1996, in
Gamma-ray Emitting AGN, eds. J. G. Kirk, M. Camenzind, C. von
Montigny, \& S. Wagner (Heidelberg), 117

\bibitem[Wehrle(1999)]{w99}
Wehrle, A. E., 1999, Astroparticle Physics, 11, 169

\bibitem[Wehrle et al.(1993)]{wuz93}
Wehrle, A. E., Unwin, S. C., Zook, A. C., et al., 1993, BAAS, 25,
1450

\bibitem[Witzel et al.(1988)]{wsj88}
Witzel, A., Schalinski, C. J., Johnston, K. J., et al., 1988, A\&A,
206, 245

\bibitem[Xu et al.(1995)]{xrp95}
Xu, W., Readhead, A. C. S., Pearson, T. J., Polatidis, A. G., \&
Wilkinson, P. N., 1995, ApJS, 99, 297

\bibitem[Zensus(1997)]{z97}
Zensus, J. A., 1997, ARA\&A, 35, 607

\bibitem[Zhang \& B\aa\aa th(1990)]{zb90}
Zhang, F. J., \& B\aa\aa th, L. B., 1990, A\&A, 236, 47

\bibitem[Zhang \& B\aa\aa th(1991)]{zb91}
Zhang, F. J., \& B\aa\aa th, L. B., 1991, MNRAS, 248, 566

\end{thebibliography}
\end{document}